\documentclass[journal=nalefd,manuscript=letter]{achemso}
\setkeys{acs}{articletitle=false}
\setkeys{acs}{maxauthors = 10}
\setkeys{acs}{etalmode = truncate}
\usepackage[version=3]{mhchem}
\usepackage[separate-uncertainty=true]{siunitx}
\usepackage{xcolor}
\usepackage{float}
\newcommand{\figurescale}{1}
\newcommand{\onecolfigurescale}{0.75}

\author{M.~Blauth}
\affiliation[WSI]
{Walter Schottky Institut and Physik Department, Technische Universit\"at M\"unchen, Am Coulombwall 4, 85748 Garching, Germany}
\alsoaffiliation{Nanosystems Initiative Munich (NIM), Schellingstr. 4, 80799 Munich, Germany}
\author{G.~Vest}
\author{S.~Loukkose~Rosemary}
\author{M.~Prechtl}
\author{O.~Hartwig}
\author{M.~J\"urgensen}
\affiliation[WSI]
{Walter Schottky Institut and Physik Department, Technische Universit\"at M\"unchen, Am Coulombwall 4, 85748 Garching, Germany}
\author{M.~Kaniber}
\affiliation[WSI]
{Walter Schottky Institut and Physik Department, Technische Universit\"at M\"unchen, Am Coulombwall 4, 85748 Garching, Germany}
\alsoaffiliation{Nanosystems Initiative Munich (NIM), Schellingstr. 4, 80799 Munich, Germany}
\author{A.~V.~Stier}
\affiliation[WSI]
{Walter Schottky Institut and Physik Department, Technische Universit\"at M\"unchen, Am Coulombwall 4, 85748 Garching, Germany}
\author{J.~J.~Finley}
\affiliation[WSI]
{Walter Schottky Institut and Physik Department, Technische Universit\"at M\"unchen, Am Coulombwall 4, 85748 Garching, Germany}
\alsoaffiliation{Nanosystems Initiative Munich (NIM), Schellingstr. 4, 80799 Munich, Germany}
\email{finley@wsi.tum.de}

\title{Ultra-compact photodetection in atomically thin MoSe$_2$}


\begin{document}

\begin{abstract}
Excitons in atomically-thin semiconductors interact very strongly with electromagnetic radiation and are necessarily close to a surface.
Here, we exploit the deep-subwavelength confinement of surface plasmon polaritons (SPPs) at the edge of a metal-insulator-metal plasmonic waveguide and their proximity of 2D excitons in an adjacent atomically thin semiconductor to build an ultra-compact photodetector.
When subject to far-field excitation we show that excitons are created throughout the dielectric gap region of our waveguide and converted to free carriers primarily at the anode of our device.
In the near-field regime, strongly confined SPPs are launched, routed and detected in a \SI{20}{\nm} narrow region at the interface between the waveguide and the monolayer semiconductor.
This leads to an ultra-compact active detector region of only \SI{\sim 0.03}{\um\squared} that absorbs \SI{86}{\percent} of the propagating energy in the SPP.
Due to the electromagnetic character of the SPPs, the spectral response is essentially identical to the far-field regime, exhibiting strong resonances close to the exciton energies.
While most of our experiments are performed on monolayer thick MoSe$_2$, the photocurrent-per-layer increases super linearly in multilayer devices due to the suppression of radiative exciton recombination.
These results demonstrate an integrated device for nanoscale routing and detection of light with the potential for on-chip integration at technologically relevant, few-nanometer length scales.

\end{abstract}

\section{Introduction}
Information data processing systems strive towards increasing signal transmission rates while reducing power consumption.
Data transfer via metallic interconnects reach very high integration densities, due to small characteristic length scales of a
few nanometers (nm), whilst suffering from Joule heating \cite{gurrum2004thermal} and parasitic capacitances that hinder further down-scaling.
Scaling of all optical data-interconnects down towards similarly small length scales represents a major challenge, although optical data transfer and processing offers very high transmission rates \cite{Heuring.1992} and exceedingly small energy-per-bit consumption \cite{Nozaki.2010}.
Conventional optical devices, such as waveguides, splitters, sources and detectors have have characteristic sizes with a lower bound defined by the diffraction limit, of order of a few hundred nanometers at optical wavelengths.\cite{Born.2016}.
This prevents the scaling of integrated optical detector sizes to the few nanometer regime.
Here, plasmonics offers new capabilities by providing deep-subwavelength confinement of the light field at optical frequencies \cite{Ozbay.2006}.

In this Letter, we demonstrate a proof-of-principle device in which electromagnetic signals are routed along plasmonic waveguides and detected in an ultra compact detector based on a monolayer of MoSe$_2$ with an active area of only \SI{\sim 0.03}{\um\squared}.
Such proof of principle devices bridge the gap between the current state-of-the-art conventional photodetectors and the technologically relevant nanometer regime.
Due to their strong light-matter interaction and easy integration into lithographically defined structures, atomically thin two-dimensional materials are ideally suited for such ultrascaled electro-optical devices.
For example, fast data processing and broadband light detection was already demonstrated using graphene \cite{schwierz2010graphene,gan2013chip}.
However, the lack of a bandgap \cite{bonaccorso2010graphene} leads to weak optically mediated changes of the free carrier density, low on-off ratios \cite{schwierz2010graphene} and limited light absorption (\SI{\sim 2.3}{\percent} per atomic layer).
On the other hand, atomically thin transition-metal dichalcogenides (TMDCs) have a direct bandgap in the visible light range \cite{Splendiani.2010,Mak.2010}, could potentially support new functionalities due to the optically accessible valley degree of freedom \cite{DiXiao.2012,Cao.2012,Mak.2012,Zeng.2012,Sallen.2012} and can have near-unity radiative internal quantum efficiencies \cite{Amani.2015}.
Recent experiments have demonstrated TMDC-based photo detection in microscopic FET structures \cite{radisavljevic2011single,wang2012electronics,lopez2013ultrasensitive,koppens2014photodetectors,wang2016review,lee2018high} as well as nanoscale systems via propagating plasmonic modes in chemically synthesized nanowires based on pick-and-place fabrication techniques \cite{goodfellow2015direct}.
Furthermore, atomically thin MoS$_2$ was incorporated into Si$_3$N$_4$ waveguides showing integrated photo detection with a photo detector footprint of a few \SI{}{\um\squared} \cite{marin2019mos}.
In our device, we reduced the active detector area by about two orders of magnitude and the total device footprint to a level approximately one order of magnitude smaller than conventional active photodetectors having the same high absorption efficiency.
Since numerical simulations show that our device can be optimized further, our results show a path towards on-chip plasmon photodetectors at the nanoscale with the possibility of dense integration.

We have realized a plasmonic system consisting of a lithographically defined waveguide and a mechanically exfoliated MoSe$_2$ monolayer, enabling interactions between the strongly absorbing TMDC flake and tightly confined plasmonic modes.
The use of electron-beam lithography provides full control over the position and geometry of the plasmonic waveguide, allowing deterministic routing and detection of photons on-chip.

Numerical calculations of the fabricated structures yield an overlap of the plasmonic mode and the TMDC monolayer of only $l_{mode} = \SI{20}{\nano\meter}$ for the detection of routed signals, corresponding to an ultra compact active detector region of $A = \SI{0.03}{\um\squared}$ (\SI{86}{\percent} absorption, equals $1/e^2$ transmission) and a total device footprint of \SI{0.3}{\micro\meter\squared}.
Electro-optical experiments demonstrate a tunable responsivity of up to $R = \SI{18}{\milli\ampere\per\watt}$ under far-field excitation.
Our results pave the way toward on-chip plasmon photodetectors at the nanoscale with the possibility of dense integration.

\section{Results and discussion}
\begin{figure}[!ht]
\scalebox{\figurescale}{\includegraphics{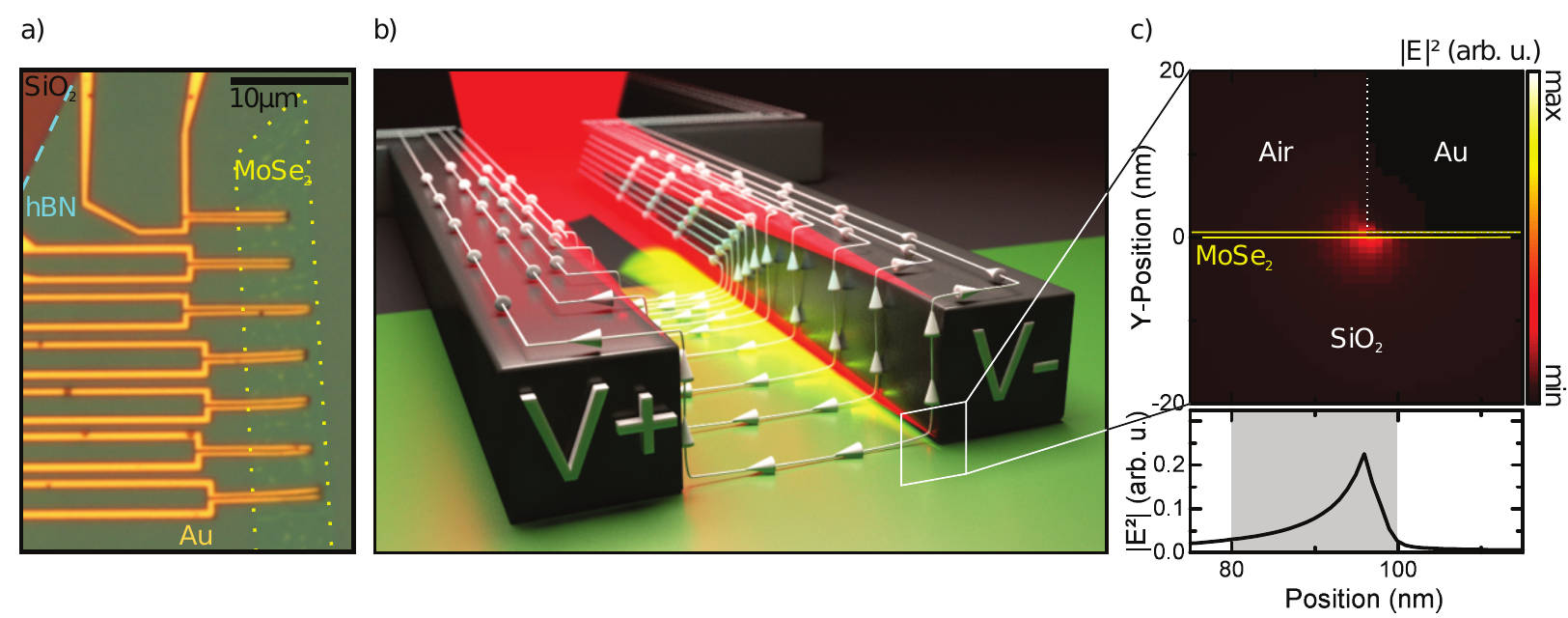}}
\renewcommand{\figurename}{Figure}
\caption{\label{fig1}
Device layout and working principle.
(a) Optical microscope image of the fabricated structures consisting of a bottom hBN multilayer (blue outline), a MoSe$_2$ monolayer (yellow outline) and gold plasmonic slot waveguides partially covering the TMDC monolayer (orange).
(b) Artists impression of the combined system consisting of a contacted plasmonic slot-waveguide and a MoSe$_2$ monolayer flake.
Depicted is the near field configuration in which the excitation laser is coupled into the waveguide at the antenna structure.
Surface-plasmon polaritons (SPPs) travel along the waveguide and are resonantly absorbed by the TMDc monolayer.
The created excitons are converted into photocurrent through the strong interfactial potential at the TMDC/waveguide interface.
(c) Squared electric field distribution of the fundamental propagating mode at the edge region (green outline in b).
The TMDC monolayer (metal slab) is marked with a solid yellow (white) outline.
The field maxima coincide with the monolayer position enabling efficient coupling of the SPPs to the monolayer.
Inset: Line cut of the field distribution along the monolayer flake.
The shaded area shows the modal area of $l_{mode} = \SI{20}{\nm}$, which contains \SI{86}{\percent} of the total SPP energy.
}
\end{figure}
The sample studied consists of several plasmonic slot waveguides which are defined by two gold metallic strips partially covering a MoSe$_2$ monolayer.
Figure~\ref{fig1}~a shows an optical microscope image of the region of interest.
A hexagonal boron nitride (hBN) multilayer (blue outline) was transferred onto a Si/SiO$_2$ substrate (red) using a dry viscoelastic stamping method \cite{CastellanosGomez.2014} to provide a controlled dielectric environment, reduced substrate surface roughness and increased adhesion for subsequent fabrication steps.
In a next step, a MoSe$_2$ monolayer (yellow outline) was transferred onto the hBN region and, finally, the plasmonic slot waveguides were fabricated using standard electron beam lithography and gold evaporation methods.
The left-hand side in figure~\ref{fig1}~a shows the contacts facilitating biasing whereas in the center, the waveguides are visible, partially covering the MoSe$_2$ monolayer.

A schematic image of our plasmonic slot waveguides is shown in Figure~\ref{fig1}~b.
Surface plasmon polaritons (SPPs) are launched by a far field excitation laser that is focused on the antenna structure (for details on allowed modes and coupling efficiencies see Supplementary Information section S1).
The structure converts the incident photons into SPPs and routes them to the detection region, where the slot waveguide is on top of the MoSe$_2$ monolayer.
There, the SPPs are absorbed and the photogenerated charges are extracted and measured across the biased metal waveguides.
In contrast to other TMDC far-field \cite{yin2011single,koppens2014photodetectors,wang2014two,zhang2016van} and waveguide-based \cite{marin2019mos} photodetectors, this geometry features a truly nanoscale detection area:
Figure~\ref{fig1}~c depicts a zoom-in of a waveguide edge, indicated by the green outline in panel~b.
The color scale indicates the magnitude of the numerically simulated \cite{LumericalSolutionsInc..} electric field distribution squared ($|E(r)|^2$) of a plasmonic mode supported by the waveguide (MoSe$_2$ refractive index data from ref. \citenum{li2014measurement}).
The modal area of the plasmonic mode, $A_{mode} = 0.016\lambda^2$, was extracted from numerically simulated mode profiles that showed deep-subwavelength confinement of the propagating SPPs.
The field maximum at the edge of the metal slab coincides with the location of the monolayer flake, thus, providing efficient coupling of the SPP to the monolayer.
To elucidate the interaction of the SPPs with the MoSe$_2$ flake, the plot in the lower panel of figure~\ref{fig1}~c shows a cross-section of $|E(r)|^2$ through the TMDC monolayer.
As can be seen by the shaded area in the lower panel, the overlap of the mode profile and the monolayer flake is localized within a lateral extent of only $l_{mode} = \SI{20}{\nm}$.
Furthermore, frequency-resolved numerical simulations yielded a group velocity of this bound mode of $v_g = 0.78 c$, emphasizing the dominating photon-like character of the SPP.
Thus, the plasmonic waveguide facilitates deep-subwavelength confinement while preserving the rapid propagation of the SPP along the waveguide.

\begin{figure}[!hp]
\scalebox{\onecolfigurescale}{\includegraphics{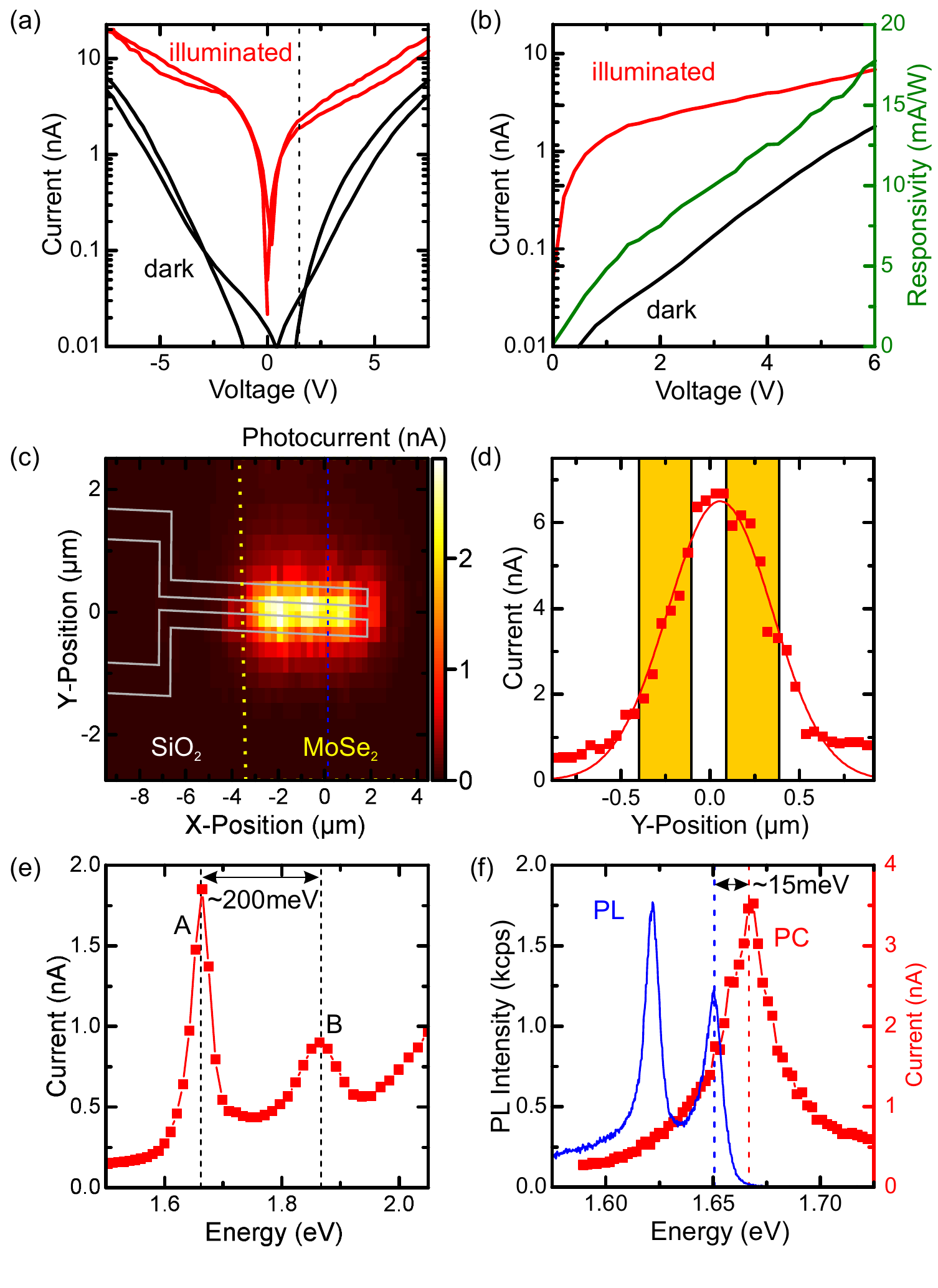}}
\renewcommand{\figurename}{Figure}
\caption{\label{fig2}
Photocurrent characteristics
(a) I(V) measurements performed on dark (illuminated) waveguide in black (red).
The dark measurement reveals high resistance region up to \SIrange{-4.2}{3}{\volt} followed by exponentially increasing tunneling currents.
Illuminated measurement shows photo-activated conductivity significantly surpassing the dark measurement.
(b) High-resolution measurement of the dark (black) and bright (red) current-voltage characteristics, respectively.
Green curve indicates responsivity computed from presented I(V) data as a function of applied extraction voltage.
(c) Spatial dependency of the photocurrent (PC) vs excitation position, waveguide (monolayer) positions indicated by the white (yellow) outlines.
(d) High-resolution spatial scan of the PC signal along the blue line in c, metal slab positions indicated by yellow rectangles.
(e) Coarse spectrally resolved PC signal exhibiting two resonances at \SI{1.663}{\electronvolt} and \SI{1.868}{\electronvolt}, respectively.
(f) Fine excitation energy dependent measurement around the A resonance observed in e.
PC data in red compares to photoluminescence data in blue recorded under CW HeNe excitation.
}
\end{figure}

In the following, we explore the photocurrent (PC) response of the device under far-field laser illumination.
Figure~\ref{fig2}~a shows I(V) measurements at $T = \SI{15}{\kelvin}$ performed by biasing an individual plasmonic slot waveguide across its two contacts.
The data indicated in black were recorded without external illumination and reveal a high-resistance region ($R > \SI{10}{\giga\ohm}$) for bias voltages in the range of \SIrange{-4.2}{3}{\volt}.
For voltages outside this window, the current increases exponentially in both biasing polarities indicative of tunneling processes.
This observation is consistent with a back-to-back Schottky diode structure formed at the interface between the metal waveguide and the TMDC monolayer (see Supplementary Information section S2).
Moreover, the asymmetries in the two bias directions are consistent with local variations of the contact resistance between the metallic waveguides and the monolayer.
Furthermore, weak hysteresis is observed, indicating the presence of charge trapping centers\cite{lee2013flexible}.
The red curve shows the measured current under continuous-wave (CW) Ti:Sapphire laser excitation at \SI{1.669}{\electronvolt} centered on the waveguide-TMDC system at a power density of \SI{120}{\watt\per\cm\squared} (\SI{1}{\uW} power on a diffraction limited spot size of \SI{1}{\um}).
Here, the extracted currents are significantly enhanced compared to the dark measurement and the absence of the high-resistance region around zero volts indicates the extraction of photo-activated charge carriers.

In Figure 2b, we determine the responsivity $R$ of the detector from the photocurrent behavior in the non-saturating bias range between \SIrange{0}{6}{\V}.
The responsivity $R = \eta_{ff} (I_{PC} - I_{dark})/P_{in}$ was extracted from the filling factor $\eta_{ff}$, the illuminated ($I_{PC}$) and dark ($I_{dark}$) current and the incident laser power ($P_{in}$).
The filling factor $\eta_{ff} = 0.29$ was calculated from the modal overlap of the Gaussian laser spot with the active region of the detector.
The responsivity increases monotonically to values up to $R = \SI{18}{\milli\ampere\per\watt}$ for voltages of $V_{bias} = \SI{6}{\volt}$ with photocurrent $I_{PC} = \SI{6.9}{\nA}$ and dark current $I_{dark} < \SI{1.8}{\nano\ampere}$.
The monotonic increase of the responsivity arises due to the larger extraction efficiency for larger bias voltages which increasingly dominates over the dark current.
The responsivity reported here is almost \SI{40}{\percent} higher than a previous finding for CVD-grown MoSe$_2$ monolayers, which reported a value of \SI{13}{\mA\per\W} \cite{xia2014cvd}.
For multilayer MoSe$_2$ photodetectors, responsivities exceeding \SI{100}{\A\per\W} have been demonstrated employing back-gating to control carrier density and enhance device performance \cite{ko2017high,lee2018high}.
We discuss multilayer response of our devices towards the end of this Letter.

To probe the spatial distribution of the photocurrent signal, the extraction voltage was fixed at $V_{bias} = \SI{1.5}{\volt}$ (responsivity $R = \SI{6.5}{\mA\per\W}$) and the waveguide was spatially scanned with the laser tuned to \SI{1.669}{\electronvolt}, just above the A-exciton resonance of MoSe$_2$, at a constant power density of \SI{360}{\watt\per\cm\squared}.
Figure~\ref{fig2}~c shows a colormap of the extracted photocurrent as a function of excitation position.
The edge of the monolayer flake is indicated by the yellow dashed line on figure~\ref{fig2}~c.
Clearly, the photocurrent is only observed in the region where the waveguide and the monolayer overlap.
Towards the edge of the flake (left-hand side of figure~\ref{fig2}~c) and towards the end of the waveguide (right-hand side) the photocurrent diminishes as expected.
To precisely locate the spatial origin of the PC signal, we recorded a high-resolution PC line scan along the dashed blue line indicated in panel~b.
The results are shown in figure~\ref{fig2}~d where the yellow rectangles mark the metal slab positions.
The solid red curve is a Gaussian fit to the PC signal with a full width at half-maximum (FWHM) of \SI{690 \pm 30}{\nano\meter}.
Comparing this to the excitation spot size (FWHM $d = \SI{608 \pm 6}{\nano\meter}$) shows that the spatial extent of the PC signal is dominated by the excitation spot size.
We conclude that the PC signal primarily originates from the \SI{\sim 200}{\nm}-wide gap region of the waveguide which thus defines the active region of the detector.
Because we used gold electrodes, the extracted photocurrent shows unipolar behavior which causes the small lateral offset of the maximum PC signal position by \SI{62 \pm 9}{\nano\meter} from the center of the waveguide (for details, see Supplementary Information section S2).

We now investigate the form of the photocurrent spectrum.
Hereby, the emission from a pulsed supercontinuum white light laser was spectrally filtered to a bandwidth of \SI{12}{\meV} and tuned between \SIrange{1.5}{2.1}{\electronvolt} at a fixed optical excitation power density of \SI{120}{\watt\per\cm\squared}.
Figure~\ref{fig2}~e shows the results:
Two distinct resonances are observed, centered at \SI{1.663}{\electronvolt} and \SI{1.868}{\electronvolt}, respectively.
The absolute position and splitting (\SI{\sim 200}{\milli\electronvolt}) matches the energy spacing between the A and B exciton reported in literature \cite{Ross.2013,wang2015exciton}.
Thus, we identify the resonances as the A and B excitons of MoSe$_2$, substantiating that the photocurrent signal is excitonic in origin.
Note that due to the large FWHM of our excitation laser (FWHM$ = \SI{12}{\milli\electronvolt}$), the linewidth of our A-exciton resonance is resolution-limited.

A high-resolution scan performed around the A exciton using a Ti:Sapphire laser (FWHM$ < \SI{100}{\micro\eV}$) produces the spectrum presented in figure~\ref{fig2}~f.
For comparison, the blue data is a photoluminescence spectrum recorded from the monolayer.
The observed small Stokes shift of \SI{\sim 15}{\milli\electronvolt} is indicative of the good optical quality of the TMDC material.
Considered together, our results are strongly indicative that, subject to far-field illumination PC arises from the generation of excitons and their ionization at the negative electrode.
The observed photocurrent spectrum is in good agreement to other reports on TMDC photoresponse in the literature \cite{klots2014probing,lopez2013ultrasensitive,hwan2014high}.

\begin{figure}[!ht]
\scalebox{\figurescale}{\includegraphics{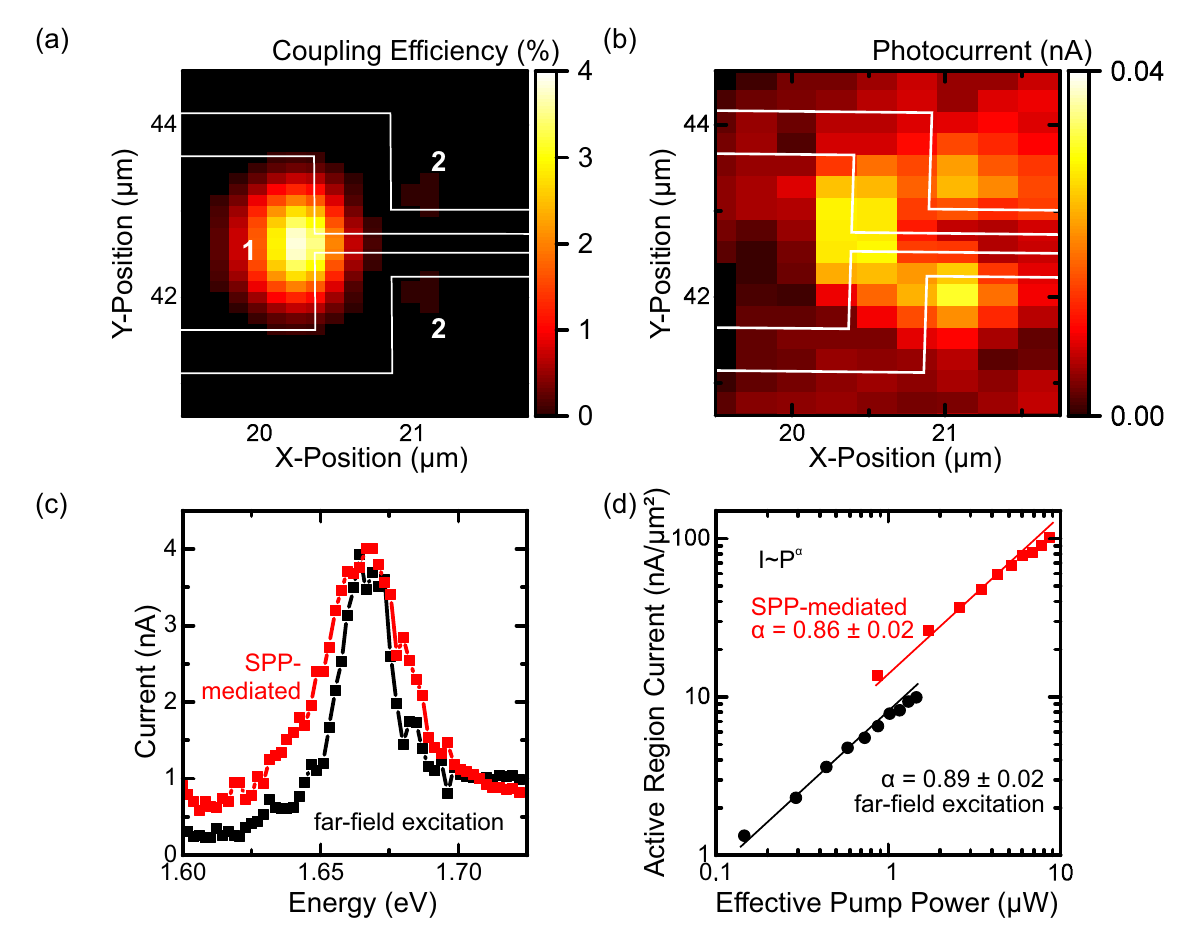}}
\renewcommand{\figurename}{Figure}
\caption{\label{fig3}
SPP-mediated photocurrent extraction
(a) Numerical simulations of the input-coupling efficiency of far-field excitation to propagating SPPs at the antenna as a function of excitation position.
(b) Photocurrent measurements as a function of excitation laser spot position around the waveguide antenna structure.
Excitation conditions are \SI{360}{\watt\per\cm\squared} at \SI{1.669}{\electronvolt}, extraction voltage is \SI{1.5}{\V}.
(c) Spectrally resolved PC resonance in far-field excitation (black) and SPP-mediated (red) geometry.
For far-field and SPP-mediated excitation the laser spot was centered on the waveguide gap along the blue line in Figure~\ref{fig2}~c and position 1, respectively.
(d) Photocurrent normalized to active detector area $A$ as a function of effective power arriving at the detector for both far-field (black, $A = \SI{0.2}{\um\squared}$) and SPP-mediated (red, $A = \SI{0.03}{\um\squared}$) excitation geometry.
}
\end{figure}
We continue to investigate the near-field excitation of the TMDC caused exclusively by SPPs.
Hereby, we present spatially, spectrally and polarization-resolved data to unambiguously prove the generation of photocurrent by SPPs.
We start with the spatial signature of SPP generation:
The input-coupling efficiency of photons to propagating SPPs at the antenna was numerically computed by finite-difference time-domain (FDTD) simulations using geometric input parameters obtained from high-resolution atomic force microscopy (AFM).
Figure~\ref{fig3}~a shows a color map of the computed coupling efficiency for the used Gaussian excitation spot as a function of its position relative to the waveguide antenna.
The simulation yields one dominant area of maximum in-coupling efficiency at the end of the antenna (labelled 1) reaching up to $\eta_c = \SI{\sim 4}{\percent}$.
In addition, two areas of enhanced coupling efficiency are symmetrically located at the outer edges of the waveguide (labelled 2).
We obtained the corresponding photocurrent map by illuminating the sample with a CW Ti:Sapphire laser resonantly tuned to the A-exciton energy at \SI{1.669}{\electronvolt} with a power density of \SI{360}{\watt\per\cm\squared}.
The laser spot was scanned over the antenna of the waveguide biased at $V_{bias} = \SI{1.5}{\V}$.
Figure~\ref{fig3}~b shows the PC measurement results:
For excitation at position 1, a photocurrent up to \SI{29}{\nA} was measured.
(We note that a position-independent average background current of \SI{<9}{\pA} attributed to stray light was subtracted from the data presented in figure~\ref{fig3}~a).
Crucially, two additional PC maxima, located at the outer edges of the waveguide (position 2), were observed.
Those areas are clearly outside the gap region where no bias is applied and, therefore, any observation of photocurrent from these areas must arise from propagating SPPs.
This finding, together with the precise spatial agreement of the extracted photocurrent with the numerical simulations of the coupling efficiency, shows that the current is generated from propagating SPPs.
In this way, the combined waveguide-TMDC system provides SPP routing and detection capabilities.

To obtain detailed insight into the SPP-generated photocurrent, numerical calculations were performed, again based on the actual device geometry, yielding an absorption length of $L_{abs} = \SI{\sim 0.7}{\um}$.
For SPP-mediated excitation, the active detector region is given by the absorption length and the lateral overlap of the SPP mode with the monolayer $l_{mode}$.
As such, \SI{86}{\percent} of the SPP are absorbed over a distance of \SI{\sim 1.4}{\um} resulting in an active photodetection area in the near-field configuration of $A = \SI{0.03}{\um\squared}$.
Note that the absorption lengths for SPPs in our device is a factor of $\sim 20$ smaller than the absorption length reported for monolayer TMDCs coupled to dielectric waveguides \cite{marin2019mos}.

Further confirmation for SPP-based photocurrent generation stems from polarization-dependent PC measurements:
Exciting at position 1, for linear excitation polarization \emph{along} the waveguide, SPP based photocurrent was observed while it was suppressed in the \emph{orthogonal} polarization direction by a factor of $\sim 6 \times$.
This polarization dependent suppression and additional details in the spatial photocurrent distribution are in good agreement with numerical simulations (see Supplementary Information section S3).

Figure~\ref{fig3}~c shows a comparison of the photocurrent spectra excited in the far field (black data) and the near-field configurations (red data, excitation at position 1), respectively.
For both configurations, the photocurrent resonance is centered around \SI{1.670}{\electronvolt}, whilst the linewidth of the near-field spectrum is larger (\SI{44}{\milli\electronvolt} c.f. \SI{25}{\milli\electronvolt}).
Because the spectral responses are virtually identical, we conclude that the photocurrent generated in the SPP-mediated geometry also stems from the generation of excitons in the monolayer region.
The difference in linewidth can be accounted for by the different generation mechanisms for the two geometries:
In far-field excitation, the photocurrent is homogeneously generated over the excitation spot size, in the gap of the waveguide.
In contrast, for SPP-mediated excitation the absorption occurs over transverse length scales of a few nanometers, probing the local interface between the TMDC and the waveguide (see figure~\ref{fig1}~c).
Thus, the absorption of the SPP predominantly takes place at the edges of the metal strips which are areas of high, spatially inhomogeneous strain in the TMDC, known to increase the linewidth of excitons in TMDC monolayers \cite{conley2013bandgap}.

We continue to investigate the power dependence of the photocurrent signal for far-field and near-field excitation.
Figure~\ref{fig3}~d shows the photocurrent normalized to the active detection region as a function of the optical excitation power.
For far-field excitation (black) the active region is governed by the gap size, yielding an active area of $A = d_{spot} \cdot w _{ar} = \SI{0.2}{\um\squared}$ (spot size $d _{spot} = \SI{1.0}{\um}$, active area width $w _{ar} = \SI{0.2}{\um}$).
The measured PC signal scales sublinearly with the incident laser power ($I_{PC} \propto P_{in}^\alpha$) with a far-field (ff) exponent of $\alpha_{ff} = \SI{0.89 \pm 0.02}{}$.
Whilst, for the far-field excitation, the incident power reaches the detector without any additional losses, the SPP-mediated measurement is subject to coupling and propagation losses obfuscating the detector performance.
To compare the performance of the detector part in both geometries, numerical calculations based on the actual device geometry were performed to compensate for losses in excitation power solely present in SPP-mediated geometry.
The simulation results show that for SPP-mediated excitation (position 1 in figure \ref{fig3} a, red data in figure~\ref{fig3}~d), the effective excitation level is reduced to \SI{<2}{\percent} of the incident laser power due to far-field coupling and propagation losses (for details, see Supplementary Information section S4).
Taking these losses into account, the responsivity for near-field (nf) excitation is measured to be $R = \SI{0.48}{\milli\ampere\per\watt}$, reduced by a factor of $\sim 14 \times$ compared to the far-field excitation.
Although the responsivity is reduced for our proof of principle device (for details see Supplemental Information Section S5), the exponent of the photocurrent is very similar in the SPP-mediated, $\alpha_{nf} = \SI{0.86 \pm 0.02}{}$, again consistent with the above picture of a very similar photocurrent generation processes.

We anticipate that significant improvements in device performance can be achieved by improving fabrication precedures and device geometry.
For example, realistic numerical simulations that include the measured waveguide morphology and roughness indicate that ohmic losses in the detector area amount to ca. \SI{18}{\percent}.
By switching to silver-based plasmonic waveguides instead of gold, further reductions in ohmic losses are expected (a detailed discussion on improving monolayer devices is given in Supplemental Information Section S5).
\begin{figure}[ht]
\scalebox{\figurescale}{\includegraphics{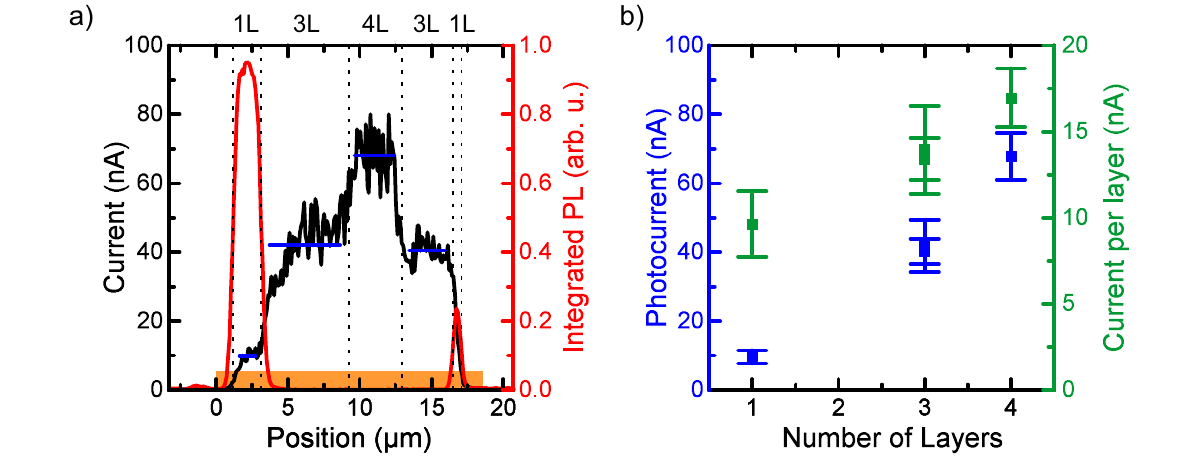}}
\renewcommand{\figurename}{Figure}
\caption{\label{fig4}
Photocurrent and photo-luminescence for different layer numbers of the MoSe$_2$ crystal.
(a) Photocurrent (black) and integrated PL (red) data for a laser spot scan along a waveguide (orange) covering a MoSe$_2$ flake.
Vertical dashed lines indicate the regions of different layer number as extracted from AFM data, blue lines show mean PC signal for each region.
Excitation conditions are \SI{700}{\watt\per\cm\squared} at \SI{1.96}{\electronvolt} (HeNe) at a bias voltage of $V_{bias} = \SI{1}{\V}$.
(b) Average photocurrent extracted from panel~a as a function of layer number (blue) and photocurrent per layer (green).
}
\end{figure}
The most significant improvement in detector performance comes however from switching from monolayer TMDC to multilayer detectors.
Figure~\ref{fig4}~a shows the photoresponse of a similar waveguide sample that was excited in the far field configuration with a MoSe$_2$ flake containing mono- and multi-layer regions with varying thickness.
For the monolayer region (1L), the simultaneously recorded PL intensity (red) is significant, in contrast to multilayer regions, that evolve towards indirect bandgap semiconductors with increasing number of layers \cite{Splendiani.2010,Mak.2010}.
Thus, photoluminescence in multilayers quickly diminished with increasing material thickness, while the excitonic absorption remains strong \cite{Mak.2010,Splendiani.2010}.
The photocurrent in the different regions of the flake is shown to discretely depend on the layer thickness, as reflected by the increased photocurrents at the corresponding positions of three-layer (3L) and four-layer (4L) thickness.
In general, an increased photocurrent is expected for thicker samples due to the increased absorption in the TMDC.
However, as indicated by the green symbols, the normalized photocurrent \emph{per-layer} increases significantly from \SI{9.6 \pm 1.9}{\nA} in the monolayer to \SI{17.0 \pm 1.7}{\nA} in the four-layer region.
This superlinearity is consistent with the absence of significant radiative recombination in thicker layers that compete with the photocurrent generation mechanism.
For multilayer MoSe$_2$, the absorption in the flake during propagation can be increased significantly, e.g., for ten layers of MoSe$_2$ the total detector footprint can be reduced to \SI{< 0.1}{\micro\meter\squared} with the same absorption properties (see Supplementary Information section S6).
Consequently, employing thicker TMDC flakes would be highly beneficial for both detector footprint as well as responsivity.

\subsection{Summary}
In summary, we have optically and electrically characterized a composite device consisting of lithographically defined plasmonic slot-waveguides and a MoSe$_2$ mono- and few-layer crystal obtaining photodetection both in far-field and near-field excitation geometry.
In far field excitation, the photoresponse stems from excitonic absorption and generation of photocurrent at the anode of our device.
In this geometry, the photocurrent is homogeneously generated over the \SI{\sim 200}{\nano\meter}-wide gap region between the metal contacts yielding an effective detector footprint of \SI{0.3}{\micro\meter\squared}.
A monotonically tunable responsivity of up to \SI{18}{\milli\ampere\per\watt} was observed.
In the near-field excitation scheme, propagating surface plasmon-polaritons have been generated and routed along the waveguide.
The detection occurs in a region of \SI{20}{\nm} in the direct vicinity of the metal slabs resulting in an active region of \SI{0.03}{\um\squared}.
The photocurrent spectrum under the near-field excitation is nearly identical to the far-field photocurrent spectrum, confirming a similar photocurrent generation mechanism.
We therefore show a path towards ultra-compact photodetection via SPP in the technologically relevant nanometer regime with potenial for dense on-chip integration.

\begin{acknowledgement}
We thank Elmar Mitterreiter and Jonas Kiemle for support in sample fabrication.
Furthermore, we gratefully acknowledge financial support by the Deutsche Forschungsgemeinschaft (DFG, German Research Foundation) under Germany's Excellence Strategy – EXC 2089/1 – 390776260, as well as support of the Technische Universit\"at M\"unchen (TUM) - Institute for Advanced Study, funded by the German Excellence Initiative and the TUM International Graduate School of Science and Engineering (IGSSE).
\end{acknowledgement}

\section{Author contributions statement}
M.K., J.J.F, M.B. and G.V. designed the study. S.L.R, M.P., O.H. and M.B. fabricated the waveguide structures, M.P. and S.L.R. exfoliated and transferred monolayers. M.J. and M.P. supported additional sample fabrication steps. M.B. built the optical setup and together with G.V. conducted optical and electrical measurements and performed the data analysis; M.P., M.J. and M.B. implemented FDE and FDTD simulations.
All authors discussed the results.
M.B., G.V., A.V.S. and J.J.F. wrote the manuscript with contributions from all other authors.
J.J.F., M.K. and A.V.S. supervised the project.

\subsection{Note}
The authors declare no competing financial interest.

\begin{suppinfo}

\end{suppinfo}

\bibliography{paper}

\end{document}